\documentclass[fleqn,usenatbib,useAMS]{mnras}

\usepackage{graphicx}	
\usepackage{amsmath}	
\usepackage{amssymb}	
\usepackage{multicol}   
\usepackage{bm}		
\usepackage{pdflscape}	

\usepackage[T1]{fontenc}
\usepackage{ae,aecompl}

\usepackage{newtxtext,newtxmath}

\title[Single Pulses from PSR B0809+74 using POLFAR]{Single Pulse Emission from PSR B0809+74 at 150 MHz using Polish LOFAR station}

\author[Basu et al.]{
Rahul Basu$^{1}$\thanks{rahulbasu.astro@gmail.com}
Wojciech Lewandowski$^{1}$
Jaros\l{}aw Kijak$^{1}$
Bartosz	\'Smierciak$^{2}$
Marian Soida$^{2}$
\newauthor
Leszek B\l{}aszkiewicz$^{3}$ 
Andrzej Krankowski$^{3}$
\\
$^{1}$Janusz Gil Institute of Astronomy, University of Zielona G\'ora, ul. 
Szafrana 2, 65-516 Zielona G\'ora, Poland.\\
$^{2}$Astronomical Observatory of the Jagiellonian University, ul. Orla 171, 
Krak\'ow, Poland.\\
$^{3}$Space Radio-Diagnostics Research Centre, University of Warmia and Mazury, 
ul. Prawochenskiego 15, 10-720 Olsztyn, Poland.}

\pubyear{2023}

\begin{document}
\label{firstpage}
\pagerange{\pageref{firstpage}--\pageref{lastpage}}
\maketitle

\begin{abstract}
We report the observations of single pulse emission from the pulsar B0809+74 
at 150 MHz using the Polish LOFAR station, PL-611. The three major phenomena of 
subpulse drifting, nulling and mode changing associated with single pulse
variations are prominently seen in these observations. The pulsar has a single 
component conal profile and the single pulses are primarily in the ``normal'' 
drift mode with periodicity ($P_3$) 11.1$\pm$0.5 $P$ for 96\% of the observing 
duration, while the shorter duration ``slow-drift'' mode has $P_3$ = 
15.7$\pm$1.2 $P$. We were able to measure the phase behaviour associated with 
drifting from the fluctuation spectral analysis that showed identical linear 
phase variations across the pulse window for both modes despite their different
periodic behaviour. Earlier studies reported that the transitions from the 
normal state to the slow-drift mode were preceded by the presence of nulling 
with typical durations of 5 to 10 periods. Our observations however seem to 
suggest that the transition to nulling follows shortly after the pulsar 
switches to the slow-drift mode and not at the boundary between the modes, with
one instance of complete absence of nulling between mode switching. In addition
we also detected a second type of short duration nulls not associated with the 
mode changing that showed quasi-periodic behaviour with periodicity, 
$P_N\sim44\pm7$. The variety of features revealed in the single pulse sequence
makes PSR B0809+74 an ideal candidate to understand the physical processes in 
the Partially Screened Gap dominated by non-dipolar magnetic fields.
\end{abstract}


\section{Introduction}
PSR B0809+74, one of the brightest pulsars seen over a wide frequency range, 
exhibits the three prominent phenomena associated with single pulse emission,
viz. subpulse drifting, nulling and mode changing, and as a result the radio 
emission from the pulsar has been widely studied \citep{VS70,TH71,LA83,vLKR02,
MR11,GJK12,HSW13}. The pulsar has a single component profile which bifurcates
at the lowest frequency range below 100 MHz. The profile morphology and line of
sight (LOS) geometry has been characterised in great detail \citep[see][for a 
summary]{RR14}, and the pulsar profile has been classified as conal single 
($S_d$), where the LOS cuts across the emission beam near its edge. The 
bifurcation at low frequencies is seen in many pulsars with S$_d$ type profiles
and is believed to be an effect of radius to frequency mapping, where the LOS 
cuts move away from the edge and towards the center of the emission beam with 
decreasing frequency \citep{ET_R83,ET_R93}. The single pulse emission shows 
prominent drift bands across the emission window and is an exemplar of the 
subpulse drifting phenomenon with multiple emission modes, along with other 
conal pulsars like PSR B0031$-$07 \citep{HTT70,VJ97,SMK05,MBT17}, PSR B0943+10 
\citep{DR01,BMR10,B18}, PSR B1819$-$22 \citep{BM18b,JCB22}, PSR B2303+30 
\citep{RWR05}, amongst others. 

The drifting behaviour is characterised by the periodicity, $P_3$, that signify
the repetition time of subpulses along any longitude of the emission window. 
There are several methods of measuring $P_3$, primarily using the mathematical 
technique of Fast Fourier transform (FFT). The most commonly used method is the
longitude resolved fluctuation spectra \citep[LRFS,][]{B73}, where FFT is 
carried out for a series of pulse intensities along each longitude, and the 
drifting appears as a peak frequency ($f_p=1/P_3$) in the fluctuation spectra, 
while the FFT phases corresponding to the peak amplitude measures the relative 
shift of the subpulses across the emission window. Other techniques also focus 
on estimating a second periodicity, $P_2$, the longitudinal separation between 
two adjacent drift bands, and represents an average estimate of the phase 
variations measured in the LRFS. These include the two-dimensional fluctuation 
spectrum \citep[2DFS,][]{ES02} and the harmonic resolved fluctuation spectra 
\citep[HRFS,][]{DR01}. The peak locations in both 2DFS and HRFS gives estimates
of $P_3$ as well as $P_2$. Other variants of these techniques are used to 
measure the temporal changes in the drifting behaviour by a continuous shift in
the FFT sequence within the observing duration, both for the LRFS \citep{BMM16}
as well as the 2DFS \citep[S2DFS,][]{SSW09}.

The single pulse emission properties of PSR B0809+74 has been studied in great
detail by \citet{vLKR02}, where a description of the drift-mode-null behaviour
is provided. The pulsar is primarily seen in the ``normal'' mode with prominent
phase-modulated drifting which switches to the ``slow-drift'' mode at certain 
instances with a longer $P_3$. It was reported that the boundary between the 
two modes are separated by nulls lasting between 5 and 10 periods. In addition,
the pulsar also exhibits short duration nulls lasting between 1-2 periods. 
Another relatively detailed study of the drifting behaviour in the normal mode 
was conducted by \citet{HSW13}, where observations were carried out spanning 
two orders of magnitude in frequency, between 40 MHz and 5 GHz. The study found
the drifting periodicity to be identical across the frequency range, but the 
phase variations of the subpulses within the emission window have detectable 
differences. At any given frequency the subpulses show non-linear phase 
variations across the emission window, i.e. the separation between two adjacent
subpulses changes as a function of the longitude range. The phase variations 
across the emission window are also frequency dependent, showing a gradual 
evolution in shape. It was noted in \citet{HSW13} that the non-linearity 
associated with the phase behaviour and its frequency evolution cannot be 
explained by the widely used model of subpulse drifting arising due to 
{\bf E}$\times${\bf B} of drift of sparking discharges in an inner vacuum gap 
above the dipolar polar cap \citep{RS75}.

In recent years an updated understanding of the origin of subpulse drifting has 
been proposed based on the Partially screened Gap model \citep[PSG,][]{GMG03}. 
It is expected that the inner vacuum gap is partially screened due to discharge
of thermally generated ions from the heated polar cap surface. On the other 
hand detailed measurements show that the magnetic field structure above the 
polar cap is dominated by the non-dipolar component \citep{GHM08,G17,SGZ17,
AM19,PM20,SG20}. The sparks in the PSG are formed in a tightly packed 
configuration, lagging behind the rotation motion of the star which gives rise 
to the observed drifting behaviour \citep{BMM20b,BMM22}. The above model has 
been successful in explaining the non-linear drifting phase behaviour in 
certain pulsars \citep{BMM23}. The non-linearity seen in the drifting phase 
behaviour of PSR B0809+74 is an ideal testbed to further investigate the 
applicability of the model of subpulse drifting from PSG. This was our initial 
motivation to carry out detailed observations of the single pulse emission from
PSR B0809+74 using the Polish LOFAR station, PL-611 \citep{BLK18}. However, in 
the process of analysing the single pulse behaviour we have found several 
previously unreported emission features associated with all three phenomena of 
mode changing, subpulse drifting and nulling. As a result we have dedicated 
this work to report the single pulse properties of this pulsar at 150 MHz 
observing frequencies, while the detailed modelling of the subpulse drifting 
behaviour using the PSG model will be addressed in a separate paper. These 
results show that individual LOFAR stations are effective instruments to study 
single pulse emission from the brighter radio pulsars.

\section{Observations and Analysis}
The observations of PSR B0809+74 were carried out using Polish LOFAR station, 
PL-611 located in \L{}azy, near Krakow, which is part of the POLFAR Consortium 
\citep{BLK16,BLK18}. We used the HBA part of the Telescope which operates 
around a center frequency of 150 MHz with a bandwidth of 80 MHz. PSR B0809+74 
forms part of a monitoring program with POLFAR, involving around 50-100 
pulsars, to study the scintillation and scattering effects of the interstellar 
medium on pulsar signal. Each pulsar is observed for a duration of 1-2 hours on
roughly a monthly cadence. In these observations the pulsar signals are 
averaged over 10 seconds and therefore not suitable for single pulse studies. 
However, these measurements revealed high detection sensitivity for the 10 
second integrated signals from PSR B0809+74, suggesting suitability of this 
object for single pulse studies with POLFAR.

The data was obtained using the LUMP software system\footnote{Developed by 
James Anderson, https://github.com/AHorneffer/lump-lofar-und-mpifr-pulsare} and
coherently dedispersed offline to split into 366 frequency channels over a 
final bandwidth of 71.5 MHz, centered at 153.8 MHz observing frequency. Each 
pulse period was divided into 1024/2048 time bins and recorded individually 
before being combined into a single pulse sequence. We conducted separate 
observations to record the single pulse emission on two separate occasions, 25 
June, 2019, for roughly 3 hours observing duration resulting in 8116 single 
pulses, and for roughly 5.5 hours duration in 2022, resulting in 15080 single 
pulses. The detection sensitivity reduced during the last third of the later 
observation limiting some of the single pulse studies to around 11900 pulses. 
Subsequent analysis involved removing RFI across time and frequency channels as
well as averaging over the bandwidth and were implemented using the PSRCHIVE
package \citep{vSDO12} and specially designed RFI removal codes for POLFAR data
analysis. Finally, single pulse stacks were produced comprising of two  
dimensional representation with the rotation longitude along the x-axis and the
pulse number in the y-axis. All subsequent single pulse analysis were carried 
out on the pulse stacks.

\section{Emission Properties}

\subsection{Mode Behaviour} \label{subsec:mode}

\begin{table}
\caption{Statistics of the two Emission Modes}
\label{tab:modestat}
\begin{tabular}{lccccc}
\hline
 Date & $N_p$ & $N_T$ & Mode & Abundance & Avg. Length \\
   &  &  &  & (\%) & ($P$) \\
\hline
 25/06/2019 & 8116 & 12 & Normal & 95.1 & 593.5 \\
   &  &  & Slow-drift & 4.9 & 33.3 \\
   &  &  &  &  &  \\
 07/10/2022 & 15080 & 12 & Normal & 96.5 & 1119.6 \\
   &  &  & Slow-drift & 3.5 & 43.8 \\
   &  &  &  &  &  \\
 Full & 23196 & 24 & Normal & 96.0 & 856.6 \\
   &  &  & Slow-drift & 4.4 & 38.6 \\
\hline
\end{tabular}
\end{table}

\begin{figure*}
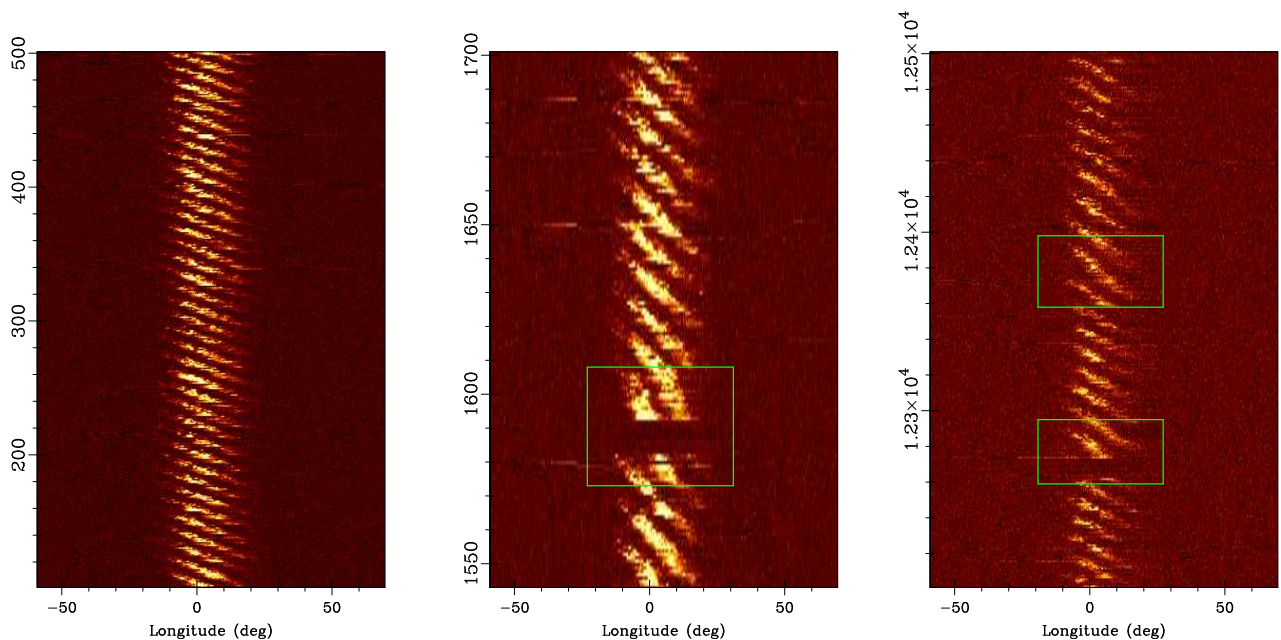

\begin{tabular}{@{}cr@{}cr@{}}
{\mbox{\includegraphics[scale=0.33,angle=0.]{B0809_sub_100-500.ps}}} &
\hspace{15px}
{\mbox{\includegraphics[scale=0.33,angle=0.]{B0809_sub_1542-1700.ps}}} &
\hspace{20px}
{\mbox{\includegraphics[scale=0.33,angle=0.]{B0809_sub_12200-12500.ps}}} \\
\end{tabular}
\caption{The figure shows three sequences of single pulse emission from PSR 
B0809+74 observed on 7 October, 2022, representing the mode changing behaviour.
The left panel shows a long sequence in the normal mode with prominent drift 
bands. The middle panel shows a transition from the normal mode to the 
slow-drift mode and back, where the box identifies the slow-drift mode. The 
switch from the normal mode to the slow-drift mode takes place around 10 
periods before the pulsar switches to the longer null state. The drift 
information appears to be preserved across the long null lasting 10 periods. 
The right panel shows two instances of the mode transitions that closely follow
each other, with the slow-drift mode once again identified within the boxes. 
The first mode transition is associated with the usual longer nulls closely 
following the transition to the slow-drift mode. However, no nulling event was 
seen during the second mode transition event.}
\label{fig:modesingl}
\end{figure*}

\begin{figure*}
\begin{tabular}{@{}cr@{}}
{\mbox{\includegraphics[scale=0.65,angle=0.]{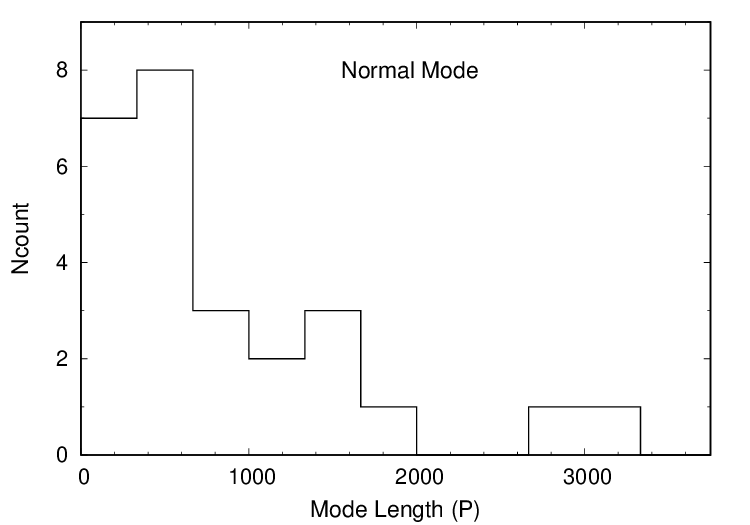}}} &
\hspace{15px}
{\mbox{\includegraphics[scale=0.65,angle=0.]{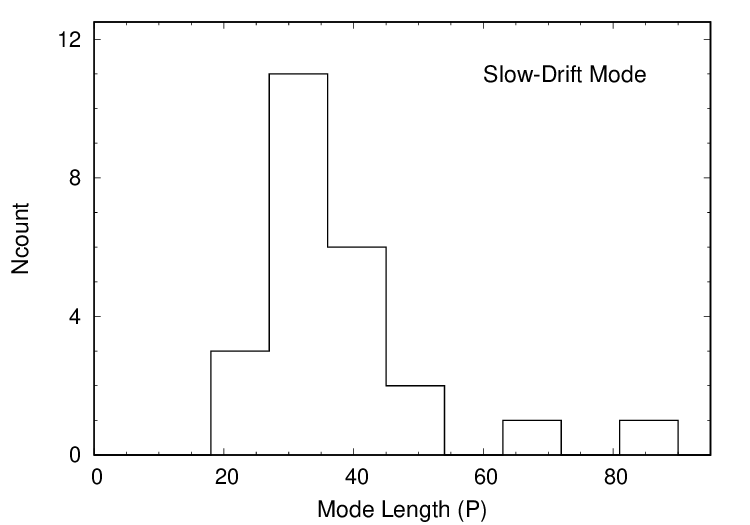}}} \\
\end{tabular}
\caption{The left panel shows the distribution of the mode lengths during the
normal state, while the right panel shows the distribution of the durations of
the slow-drift mode. Both distributions have been estimated by combining the 
observations on 25 June, 2019 and 7 October, 2022.}
\label{fig:modelen}
\end{figure*}

The detailed mode behaviour of PSR B0809+74 has been known primarily from the 
studies reported in \citet{vLKR02}. The authors used the Westerbork Synthesis 
Radio Telescope to observe the pulsar over a 18 month period, with around 
3.6$\times10^4$ measured single pulses. But individual observing sessions were 
short lived and lasted less than 1000 pulses at a time and no detailed 
measurements of mode durations were reported. The pulsar was primarily found
in the normal mode with prominent subpulse drifting. It was reported that slow 
drifting sequences start after the pulsar switches to nulls. In addition, the 
authors found two occasions, in 1999 and 2000, where the slow-drift mode lasted
for longer durations of around 120 pulses.

The longer duration observations enabled us to study the mode statistics in 
more detail. The two emission modes could be clearly identified by visual 
inspection of the pulse stacks from the two observing sessions. A detailed 
breakdown of the individual mode sequence during each observing session is 
reported in Appendix \ref{sec:app} (see Table \ref{tab:modelist}). 
Fig.~\ref{fig:modesingl} shows examples of the single pulse behaviour during 
the dominant normal mode (left panel) and transitions to the slow-drift mode 
(middle and right panels). The normal mode showed prominent drift bands at 150 
MHz similar to higher frequency behaviour. The single pulses in the slow-drift 
mode were comparatively brighter than the normal mode with much straighter 
drift bands. We found that the slow-drift mode is associated with nulling 
lasting between 3 and 11 periods at a time, as reported earlier. However, 
nulling is usually seen a few periods after the pulsar switches to the 
slow-drift mode (see Fig.~\ref{fig:modesingl}, middle panel) and does not form 
the boundary between the two emission modes, contrary to previous reports. In 
one instance nulling was entirely absent when the pulsar emission switched from
the normal mode to slow-drift mode and back (see Fig.~\ref{fig:modesingl}, 
right panel).

There were 12 transitions between the emission modes observed on each of the
observing sessions. The average statistics of the modes are summarized in 
Table.\ref{tab:modestat}. The normal mode was dominant and seen for 96\% of the
observing duration. The mode duration distribution is shown in 
Fig.~\ref{fig:modelen} (left panel), and had large variations between 31 
periods and 3312 periods, with typical values between 500-600$P$. We did not
detect the longer duration slow-drift modes lasting more than 100$P$, reported 
in \citet{vLKR02}. The distribution of the mode lengths is also shown in 
Fig.~\ref{fig:modelen} (right panel), and had relatively less spread varying 
between 21 periods and 85 periods. There were only two instances when the mode 
length exceeded 50$P$ and the slow-drift mode had typical durations lasting 
30-40$P$. The longer duration nulls seen within the slow-drift mode usually 
lasted between 3 and 11$P$ (see Table \ref{tab:modelist}). The slow-drift mode 
was seen for short durations of 5-12$P$ before the onset of the nulls and could
be identified due to their increased intensities (see Fig.~\ref{fig:modefold}) 
and change in the drift behaviour. The mode changing in this pulsar seems to 
show certain evolution with time, with the presence of longer duration 
slow-drift mode reported in earlier studies, as well as the slow-drift mode 
being more frequent during the 2019 observations compared to the 2022 
observations. However, characterising the details of the time evolution of the 
mode changing behaviour will require regular, more evenly spaced observations.

\subsection*{Average profile}

\begin{table}
\caption{Average Profile Properties}
\label{tab:modeprof}
\begin{tabular}{lc@{\hspace{1\tabcolsep}}c@{\hspace{1\tabcolsep}}c@{\hspace{1\tabcolsep}}c@{\hspace{1\tabcolsep}}c}
\hline
    & $W_{50}$ & $W_{10}$ & $W_{5\sigma}$ & \multicolumn{2}{c}{Int. Ratio} \\
   & ($\degr$) & ($\degr$) & ($\degr$) & Peak Int. & Avg. Int. \\
\hline
  Normal & 17.9$\pm$0.9 & 35.9$\pm$0.9 & 88.6$\pm$0.9 & 1.0 & 1.0 \\
   &  &  &  &  &  \\
  Slow-Drift & 16.9$\pm$0.9 & 34.8$\pm$0.9 & 40.7$\pm$0.9 & 1.34$\pm$0.01 & 1.29$\pm$0.01 \\
   &  &  &  &  &  \\
  Full & 17.9$\pm$0.9 & 35.9$\pm$0.9 & 89.6$\pm$0.9 & --- & --- \\
\hline
\end{tabular}
\end{table}

\begin{figure*}
\begin{tabular}{@{}cr@{}}
{\mbox{\includegraphics[scale=0.65,angle=0.]{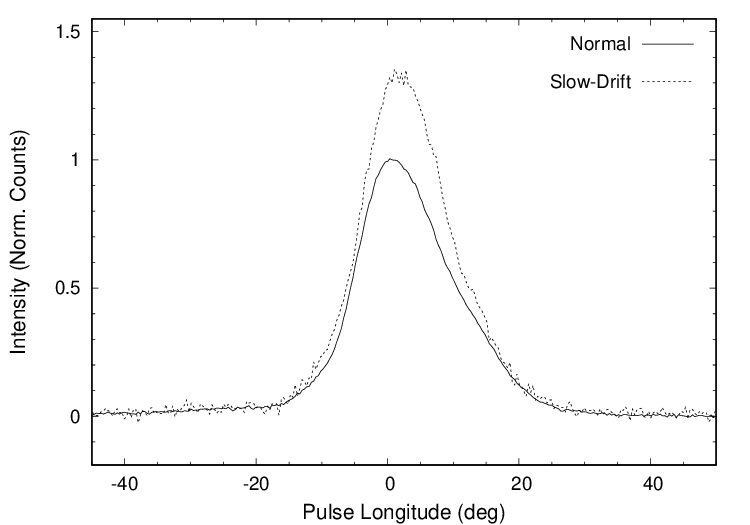}}} &
\hspace{15px}
{\mbox{\includegraphics[scale=0.65,angle=0.]{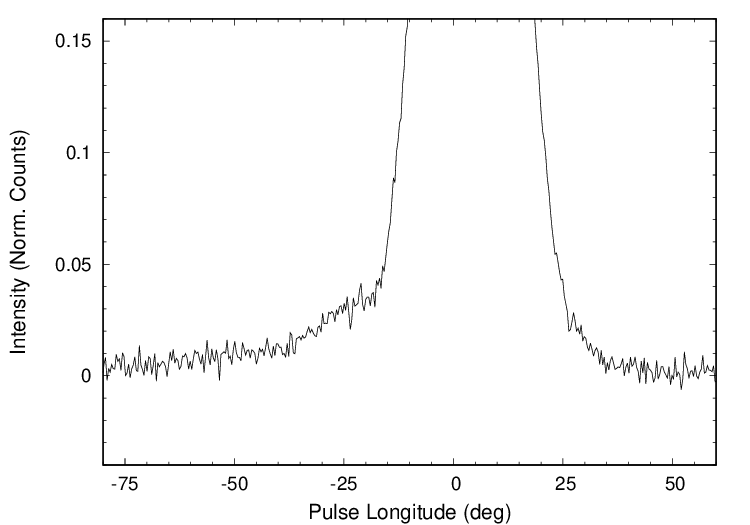}}} \\
\end{tabular}
\caption{The left panel shows the folded profiles of the normal mode and the 
slow-drift mode. The right panel shows the zoomed baseline of the average 
profile showing low level emission at the leading edge.}
\label{fig:modefold}
\end{figure*}

The average profile of the pulsar shows a single component typical of the conal
single classification \citep{ET_R93}, but exhibits an asymmetric nature with 
absorption like feature towards the trailing side which has also prompted the 
profile to be termed a `partial cone' \citep{MR11}. The average profiles of the
two emission modes are shown in Fig.~\ref{fig:modefold}, left panel, where the 
normal mode has the asymmetric nature, while the slow-drift mode is more 
symmetric and seen prominently towards the trailing side as well. The average 
profile properties are reported in Table.\ref{tab:modeprof}, including the 
widths $W_{50}$ and $W_{10}$, measured at 50\% and 10\% of the peak intensity 
level, respectively, and $W_{5\sigma}$ which specifies the emission window by 
estimating the widths above 5 times the baseline noise rms levels. The relative 
intensities of the profiles shows that the slow-drift mode is brighter than the
normal mode, with the peak intensity being around 35\% higher in the slow-drift
mode compared to the normal mode, while the average energy of slow-drift mode 
being around 30\% brighter. 

The $W_{50}$ and $W_{10}$ measurements in the two modes match within errors 
which suggests that the location of the emission region remains largely 
unchanged during mode changing \citep{BM18b,BPM19,RBMM21}. The right panel of 
Fig.~\ref{fig:modefold} shows the zoomed in baseline region of the normal mode 
profile and shows low level emission at the trailing side of the profile, 
extending from around $-40\degr$~till $-15\degr$ from the profile peak. This is
further evidenced by the large value of $W_{5\sigma}$ for this profile, which 
is more than twice the $W_{10}$ value. The extended feature is not visible in 
the slow-drift mode profile due to higher baseline noise levels. A second 
component in the pulsar profile becomes prominent at frequencies below 100 MHz 
\citep{HSH12} and it is one possibility that the component starts becoming 
visible at low levels from 150 MHz. However, in these earlier works the pulsar 
profile shows a wide evolution with frequency where a second component becomes 
visible at frequencies above 1 GHz in addition to the bifurcation of the 
profile below 100 MHz. It has been suggested that at low frequencies the second
component appears at the trailing side while the high frequency second 
component is at the leading side, and can be considered a distinct third 
component feature. Taking into account this interpretation the low level 
emission at the leading edge can very well be the remnants of the second 
(third) component from the high frequency profiles.

\subsection{Subpulse drifting}

\begin{figure*}
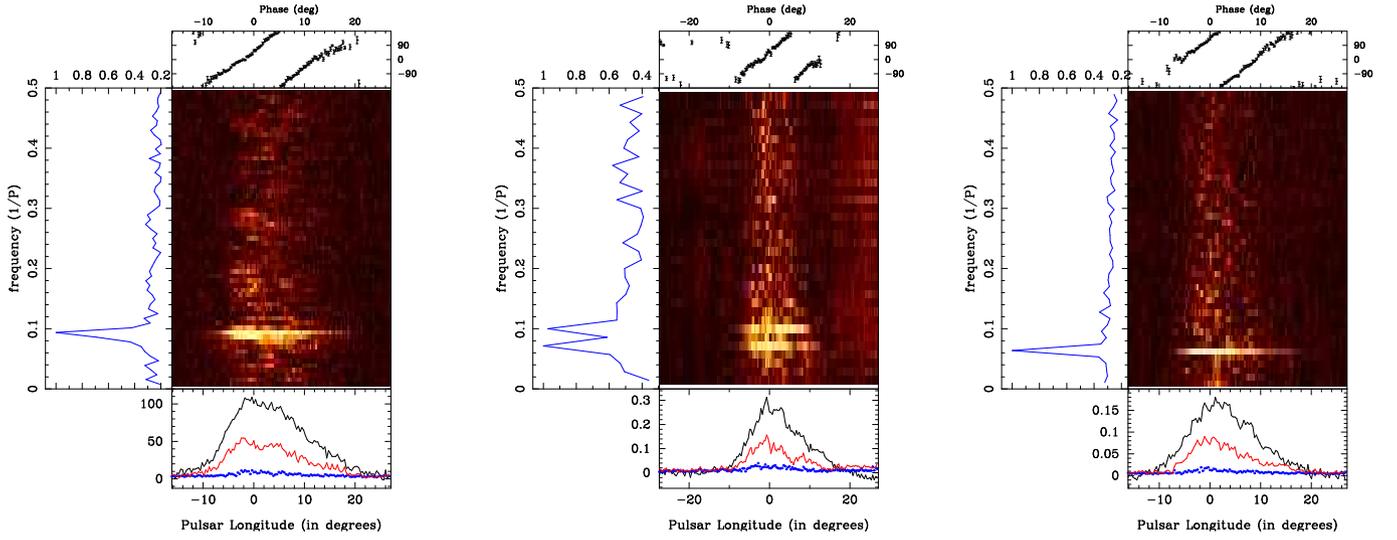

\begin{tabular}{@{}cr@{}cr@{}}
{\mbox{\includegraphics[scale=0.28,angle=0.]{B0809_ModeA_LRFS.ps}}} &
\hspace{15px}
{\mbox{\includegraphics[scale=0.28,angle=0.]{B0809_DualMode_LRFS.ps}}} &
\hspace{20px}
{\mbox{\includegraphics[scale=0.28,angle=0.]{B0809_ModeB_LRFS.ps}}} \\
\end{tabular}
\caption{The figure shows the Longitude Resolved Fluctuation Spectra (LRFS) 
from three different single pulse sequences showing the drifting behaviour in 
the two emission modes of PSR B0809+74 at 150 MHz. The left panel shows the 
LRFS of the normal mode with one clear frequency peak due to the prominent 
subpulse drifting behaviour. The central panel corresponds to a pulse sequence 
containing both emission modes and shows the two peaks, the lower peak of the 
slow-drift mode and the higher peak from the normal mode. The right panel shows
the LRFS for a sequence in the slow-drift mode.}
\label{fig:modelrfs}
\end{figure*}

\begin{table}
\caption{Subpulse Drifting Measurements}
\label{tab:modedrift}
\begin{tabular}{lcccc}
\hline
    & $f_p$ & $FWHM$ & $P_3$ & Phase Slope \\
   & ($cy/P$) & ($cy/P$) & ($P$) & ($\degr$/$\degr$) \\
\hline
  Normal & 0.090$\pm$0.004 & 0.0093 & 11.1$\pm$0.5 & 23.19$\pm$0.06 \\
   &  &  &  &  \\
  Slow-Drift &  0.064$\pm$0.005 & 0.012 & 15.7$\pm$1.25 & 25.1$\pm$0.2 \\
   &  &  &  &  \\
\hline
\end{tabular}
\end{table}

\begin{figure}
\includegraphics[width=\columnwidth]{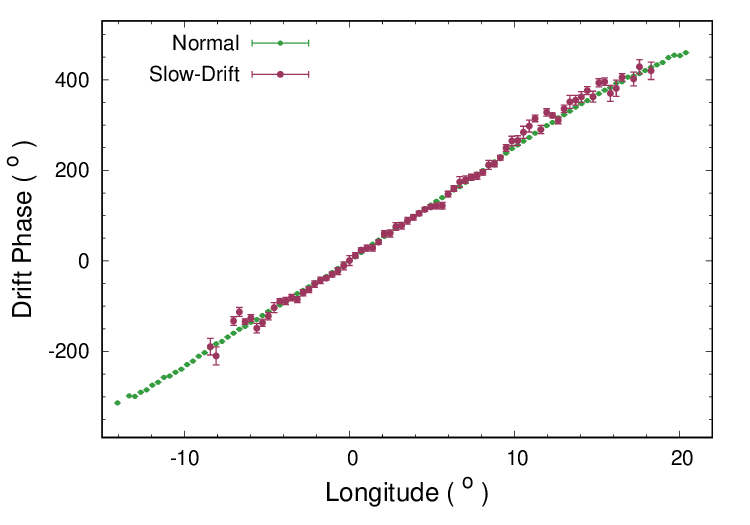}
\caption{The phase behaviour associated with subpulse drifting in the two 
emission modes. The average phase behaviour of the normal mode is shown from
the entire duration of the observation, while the phase behaviour of the 
slow-drift mode corresponds to one instance, between pulse number 12355 and 
12440 on 7 October, 2022, where these measurements were possible.}
\label{fig:driftphs}
\end{figure}

The two emission modes show visible drift bands that were studied using the
LRFS. Fig.~\ref{fig:driftphs} shows the LRFS of three pulse sequences observed 
on 7 October, 2022, corresponding to the normal mode (left panel), an 
overlapping region with both modes present (middle panel) and the slow-drifting
mode (right panel). The different drifting behaviour in the two modes is 
highlighted by their different periodic behaviour, with clearly separated peak 
frequencies seen in the middle panel of the figure. The lower frequency 
represents the longer drifting periodicity of the slow-drift mode while the 
higher frequency feature signify the periodic behaviour of the normal mode. The
normal mode usually has longer durations, lasting for several hundred periods 
at a time, and the average fluctuation spectra for all such sections was 
estimated to find the drifting periodicity as well as the phase behaviour. The 
average behaviour was obtained by estimating the LRFS corresponding to 256 
consecutive pulses at a time and subsequently shifting the starting point by 50
periods \citep[see][for details]{BMM16,BM18a}. On the other hand clear 
measurements of drifting behaviour, particularly the phase variations, were 
only possible in the slow-drift mode where the two longer sequences exceeded 50
pulses. 

Table \ref{tab:modedrift} reports the drifting measurements from the two 
emission modes, that includes the frequency peak, $f_p$, of drifting in the 
LRFS, the full width at half maximum ($FWHM$) of the frequency feature, the 
drifting periodicity, $P_3$, and the slope of the drifting phase variations. 
The $P_3$ measured from the LRFS is 11.1$\pm$0.5$P$ in the normal mode and 
15.7$\pm$1.25 $P$, which agree with earlier estimates \citep{vLKR02}. 
Fig.~\ref{fig:driftphs} shows the phase variations across the emission window 
associated with subpulse drifting in the two emission modes. The phase 
behaviour of the normal mode has been studied in earlier works \citep{HSW13}, 
and we find a mostly linear phase variation from the leading to the trailing 
edge of the emission window with a positive slope of 23.19$\pm$0.06 
$\degr/\degr$~from the average LRFS estimate (see green points in 
Fig.~\ref{fig:driftphs}). In contrast the drift phase variations of the 
slow-drift mode has not been reported in earlier works. Our estimates show (red
points in Fig.~\ref{fig:driftphs}) that the phase behaviour in slow-drift mode  
closely follows the linear phase behaviour of the normal mode with positive 
slope of 25.1$\pm$0.2 $\degr/\degr$, with possible small deviations near the 
trailing side of the profile, despite the $P_3$ in this mode being almost 50\%
longer than the normal mode.

\subsection{Nulling}

\begin{table}
\caption{Nulling characteristics}
\label{tab:Nullstat}
\begin{tabular}{ccccccc}
\hline
   & $N_T$ & $\langle BL\rangle$ & $\langle NL\rangle$ & $NF$ & $\eta$ & $P_N$ \\
   &  & ($P$) & ($P$) & (\%) &  & ($P$) \\
\hline
   &  &  &  &  &  &  \\
 All & 147 & 130.0 & 2.1 & 1.56$\pm$0.09 & 928.8 & 44.3$\pm$7.0 \\
   &  &  &  &  &  &  \\
 Mode & 20 & --- & 6.5 & 0.62$\pm$0.06 & --- & --- \\
   &  &  &  &  &  &  \\
 Short & 127 & --- & 1.5 & 0.94$\pm$0.07 & --- & --- \\
   &  &  &  &  &  &  \\
\hline
\end{tabular}
\end{table}

\begin{figure*}
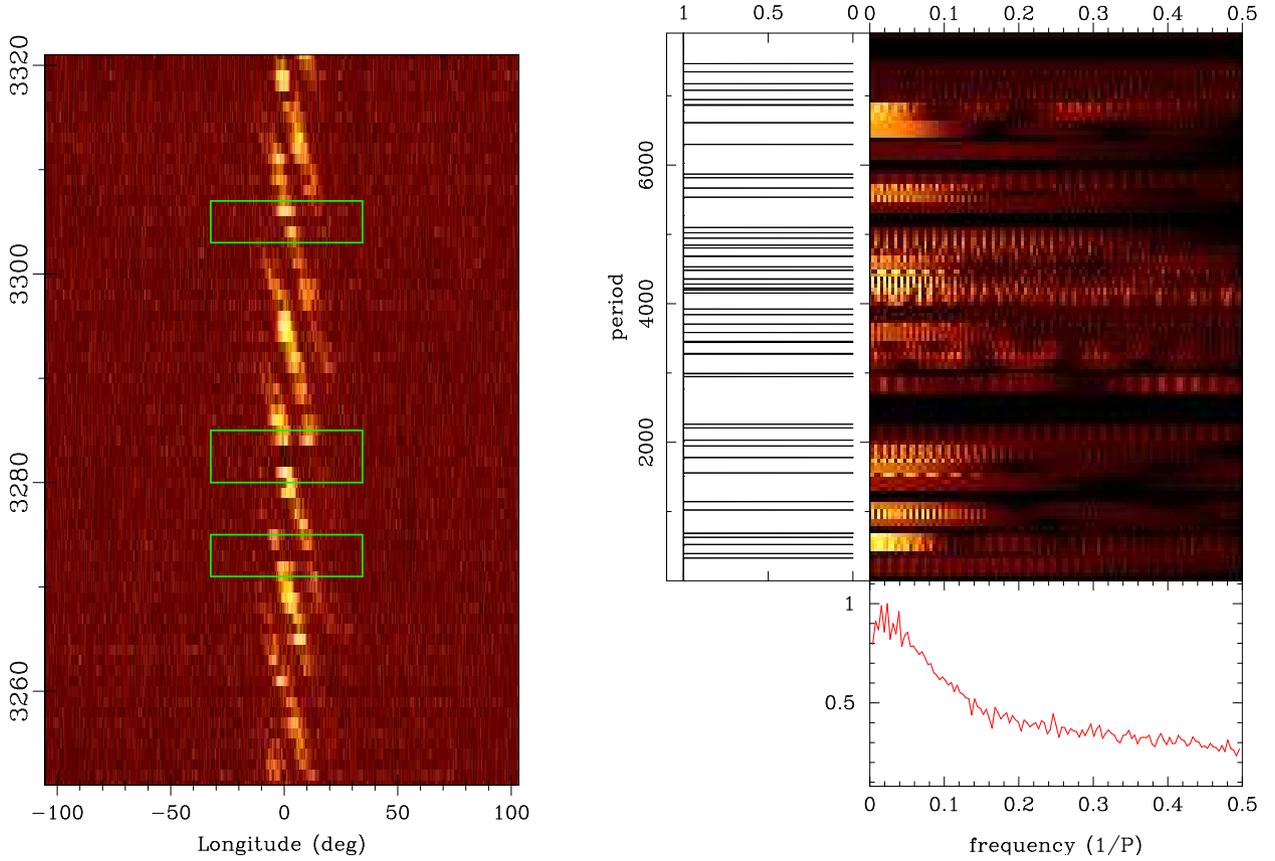

\begin{tabular}{@{}cr@{}}
{\mbox{\includegraphics[scale=0.45,angle=0.]{B0809_null_3250-3320.ps}}} &
\hspace{15px}
{\mbox{\includegraphics[scale=0.45,angle=0.]{B0809_25jul19_150MHz_nullfft.ps}}} \\
\end{tabular}
\caption{The left panel shows three examples of short duration nulling (shown 
in boxes) seen in short succession, in the single pulse sequence observed on 25 
June, 2019, at pulse number 3273, pulse number 3282 and 3283, and pulse number 
3305. The right panel shows the Fourier transform (FFT) analysis of the binary 
sequence where the nulls are identified as `0' and bursts as `1'. This binary 
series shows a wide, low frequency peak in the FFT, signifying the periodic 
nulling behaviour in PSR B0809+74.}
\label{fig:nullper}
\end{figure*}

\begin{figure*}
\begin{tabular}{@{}cr@{}}
{\mbox{\includegraphics[scale=0.65,angle=0.]{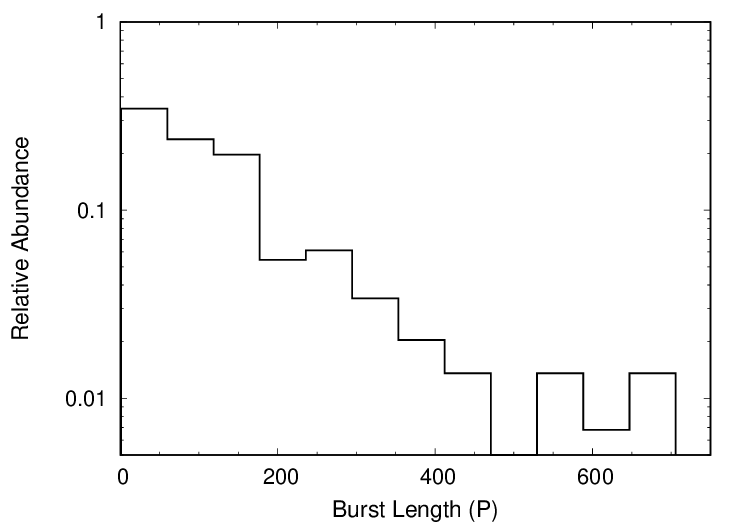}}} &
\hspace{15px}
{\mbox{\includegraphics[scale=0.65,angle=0.]{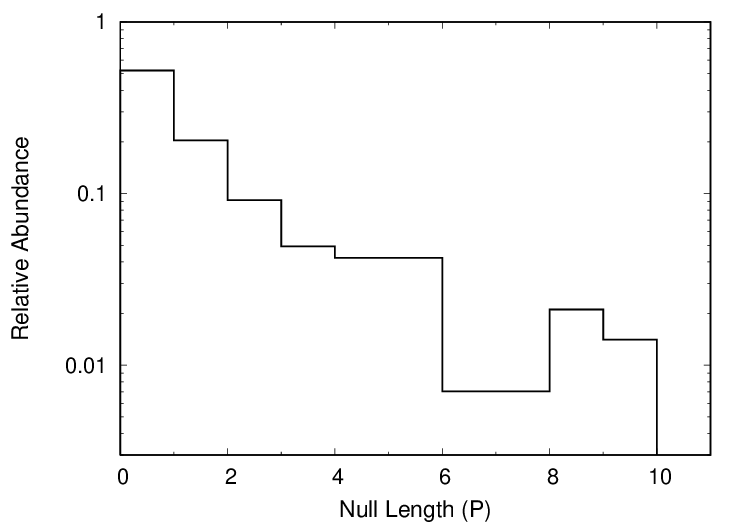}}} \\
\end{tabular}
\caption{The left panel shows the distribution of the duration of the burst
sequences obtained from both observing sessions. The right panel shows the 
corresponding distribution for the null sequence.}
\label{fig:nulllen}
\end{figure*}

\begin{figure}
\includegraphics[width=1.1\columnwidth]{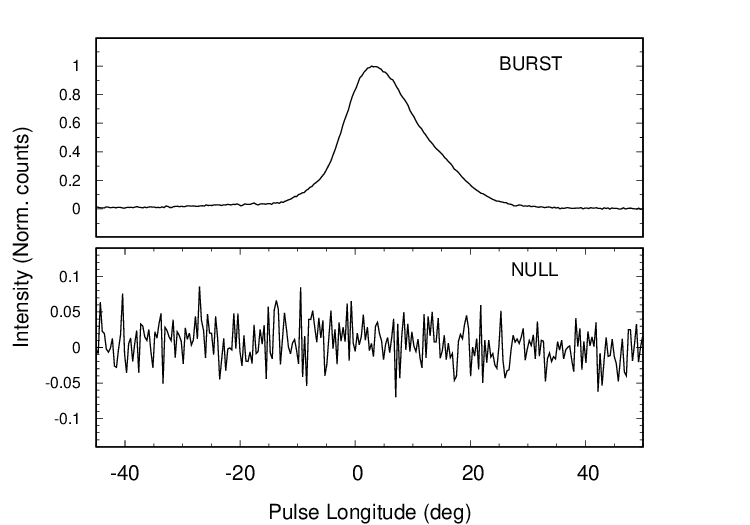}
\caption{The figure shows the average profiles obtained from the Burst and Null
pulses from the observations on 25 June, 2019.}
\label{fig:nullfold}
\end{figure}

The pulsar emission shows the presence of longer duration nulls, lasting 
between 3 and 11 periods at a time, shortly after the pulsar transitions to the 
slow-drift mode (see section \ref{subsec:mode} and also Table 
\ref{tab:modelist}). In addition we have also found the single pulse sequences 
on both observing sessions to show frequent transition to short duration nulls 
usually lasting one or two periods at a time. Hence, we report the presence of 
two distinct category of nulling behaviour in PSR B0809+74. 
Fig.~\ref{fig:nullper} (left panel) shows an example of the short duration 
nulls with three instances of transition from null to the burst state at short 
succession. We have carried out detailed analysis to characterise the nature of
the two different nulling behaviour seen in the single pulse sequence. The 
nulling measurements are usually carried out by finding statistical cutoffs 
between the null and the burst pulses and therefore require the single pulse 
emission to be measured with high detection sensitivity. The single pulse 
detection sensitivity degraded towards the last third of the observing session 
on 7 October, 2022, and we used the first 11900 pulses for the nulling 
analysis. However, this did not affect the visual identification of the 
emission modes in this duration. No such issue was encountered during the first
observing session on 25 June, 2019.

The nulling behaviour in PSR B0809+74 is reported in Table \ref{tab:Nullstat},
where the measurements from both observing sessions have been combined. The
nulling fraction ($NF$) measures the percentage of time the emission is in the
null state and was estimated to be 1.56$\pm$0.09\%, which agrees with previous
measurements \citep{vLKR02,GJK12}. Within the measurement window covering both
observing sessions we found a total of 147 transitions from the null to the 
burst state and back. Out of these, the majority of nulling, around 85\% (127), 
were short lived usually lasting 1 or 2 periods, while around 15\% (20) of 
nulling events were of longer duration, lasting between 3 and 11 periods, and 
were linked with the mode changing behaviour. The null length and burst length 
distributions are shown in Fig.~\ref{fig:nulllen}, which further highlight the 
preponderance of the short duration nulls. The average length of the short 
nulls ($\langle NL\rangle$) was around 1.5$P$ while the longer nulls associated
with the slow-drift mode had average duration of 6.5$P$. However, due to the 
relatively longer lengths the mode associated nulls made up around 40\% ($NF = 
0.62\pm0.06$ \%) of the total null pulses while the short duration nulls 
comprised the remaining 60\% ($NF = 0.94\pm0.07$ \%). The burst lengths showed 
wide variations ranging from few tens of periods (see Fig.~\ref{fig:nullper}, 
left panel) to several hundred periods at a time, with a mean length, $\langle 
BL\rangle$ = 130$P$.

The null pulses were averaged to form the null profile in 
Fig.~\ref{fig:nullfold}, lower panel, that has noise like behaviour in the
pulse window, indicating the absence of any low level emission during nulling. 
The degree of decline of the radio emission during the null state, $\eta$, can 
be obtained from the null and the burst profiles (Fig.~\ref{fig:nullfold}, 
upper panel) as $\eta = \Sigma P(i)$/3$\sigma_N$. Here, $P(i)$ is the intensity
at the $i^{th}$ longitude bin of the burst profile window, while $\sigma_N$ is 
the noise rms in the emission window of the null profile. We estimated $\eta$ = 
928.8, which is an improvement of more than a factor of five higher than 
previously reported \citep{GJK12}, and highlights that there is significant 
reduction in the emission level during nulling.

Finally, we have also studied the presence of periodic modulations associated
with the nulling behaviour. We followed the analysis scheme developed by 
\citet{BMM17}, where the single pulse sequence was converted into a binary 
series with the null pulses identified as `0' and the burst pulses as `1'. A
time varying Fourier transform (FFT) of this series was carried out for 256 
pulses at a time and subsequently shifting the starting point by 50 periods.
The FFT from the null-burst time sequence for the observations on 25 June, 2019
is shown in Fig.~\ref{fig:nullper}, right panel, and has a wide low frequency
feature. This demonstrates that in addition to subpulse drifting, PSR B0809+74 
also shows periodic nulling with periodicity, $P_N = 44.3\pm7.0P$. The 
periodic nulling behaviour is usually hidden in the LRFS due to the strong 
drifting feature and is clearly revealed in the binary sequence, where the
drifting information is suppressed. The periodic nulling behaviour is primarily 
a feature of the short duration nulls which dominate the null to burst 
transitions.

\section{Discussion}

In this work we have presented the results of the single pulse observations of 
PSR B0809+74 conducted with the Polish LOFAR station, PL-611, located near 
Krakow. Our results confirm that individual stations of the LOFAR telescope are
capable of detecting single pulse with high sensitivity, at least from several 
of the brighter pulsars. It is worth noting that the PL-611 station operates 
in the so called ``remote station'' configuration, where only a half of the 
usual international station's HBA antennas (48 instead of 96) is present, but 
is still sufficiently sensitive for the single pulse studies.

PSR B0809+74 is an well known example showing the three distinct phenomena of 
subpulse drifting, nulling and mode changing in the pulse sequence. In 
addition, we have also detected periodic nulling which is a form of periodic 
modulation, a distinct emission feature seen in around 30 pulsars 
\citep{BMM20a}. The pulsar joins a select group of sources where the periodic 
nulling coexist with subpulse drifting as well as multiple emission modes in 
the pulse sequence, with other prominent examples being PSR B0031$-$07 
\citep{BMM17}, PSR B1819$-$22 \citep{BM18b}, PSR B1918+19 \citep{HR09}, PSR 
B2000+40 \citep{BLK20}, PSR B2003$-$08 \citep{BPM19}, PSR B2319+60 
\citep{RBMM21}, etc. PSR B0809+74 is further distinguished by the presence of 
two different types of nulls, the longer duration nulls associated with the 
slow-drift mode, and more frequent short-duration nulls that contribute to 
periodic behaviour.

The physical mechanism governing the frequent transitions between the emission 
modes and the periodic nulling behaviour is not well understood. There have 
been suggestions that the change in the underlying magnetic field 
configurations near the polar cap region can modify the plasma generation 
process, leading to variations in the emission properties. A possible mechanism
for such changes in the form of perturbations introduced to the polar cap 
magnetic field by the Hall drift and thermoelectrically driven magnetic field 
oscillations has been suggested \citep{GBM21}. But more work is required to 
understand the nature of changes in the surface field due to such effects as 
well as the corresponding effect on the radio emission, before they can be used
to explain the observed behaviour.

The underlying mechanism of the subpulse drifting behaviour on the other hand 
have been studied in more detail. The plasma responsible for the radio emission
in pulsars is expected to arise due to sparking discharges from 
electron-positron pair production in an inner acceleration region (IAR) above 
the polar cap, which was initially modelled as a vacuum gap \citep{RS75}. 
During the sparking process the plasma lag behind the rotation of the star due 
to {\bf E}x{\bf B} drift in the IAR. The spark generated plasma subsequently 
populate the open field line region and give rise to the radio emission 
\citep{MGP00,GLM04}, where the lagging behind rotation is imprinted and 
observed as subpulse drifting. Several attempts have been made to understand 
the drifting behaviour in pulsars based on the vacuum gap model, where the 
magnetic field is considered to be dipolar in nature in the polar cap region, 
and the sparks are expected to rotate around the magnetic axis in the form of a
carousel \citep{GS00}. The non-linear behaviour of the drifting phase 
variations across the emission window and their frequency evolution has been 
particularly challenging to address using the above mechanism. The geometric 
model of \citet{ES02} can explain the non-linear behaviour in certain pulsars 
with deviations from linearity near the profile edges. However, an increasing 
number of pulsars have been found to show complex phase behaviour like 
bi-drifting, where different components of the profile show reversal in the 
direction of phase variations \citep{CLM05,W16,SvL17,BM18a,BPM19,SvLW20,SBD22},
large phase jumps between adjacent components \citep{BMM16,BMM19}, etc., that 
are difficult to understand using this model. Several complex scenarios like 
the presence of two different carousel motions with frequency dependent delays 
\citep{HSW13,RR14}, tilted elliptical beams \citep{WW17}, etc., have been 
proposed to explain some of these observed behaviour. Although, it is unclear 
what the physical origin of such configurations are within a vacuum gap model. 

An updated model of subpulse drifting has been proposed recently based on the 
Partially screened gap (PSG) model of the IAR, where the backstreaming 
electrons heat the polar cap surface to high temperatures, such that positively
charges ions can flow freely from the surface to partially screen the potential
difference along the IAR \citep{GMG03}. The polar cap region in a PSG is 
expected to be dominated by highly non-dipolar magnetic fields while the 
emission region has dipolar magnetic fields \citep{GMM02}. In such a 
configuration the sparks lag behind the rotation motion of the pulsar and are
constrained by the open field line boundary to exhibit two distinct drift 
patterns, along the clockwise and counter-clockwise directions around a 
stationary central spark, in the two halves of the polar cap \citep{BMM20b,
BMM22}. The phase behaviour of subpulse drifting traces the evolution of the 
sparking pattern along the line of sight (LOS) of the observer. Since the LOS 
cuts across the emission beam, the non-linear phase behaviour of subpulse 
drifting naturally arises as the magnetic field lines transition from the 
non-dipolar fields in the IAR to the dipolar fields higher up in the emission 
region. The above model has been used to find appropriate orientation of the 
surface magnetic field structure and reproduce the observed non-linear drifting
phase variations in two pulsars, PSR J1034$-$3224 and PSR B1717$-$29 
\citep{BMM23}. We intend to extend these studies to the drifting behaviour of 
PSR B0809+74~in a future work. 

The drifting phase behaviour is expected to change as a function of frequency 
due to the effect of radius to frequency mapping, where the lower frequencies 
arise from higher up the open dipolar field line region \citep{KG03}. Further 
constraints are provided by the variations of the drifting behaviour during 
mode changing. The slow-drift mode has a periodicity which is around 50\% 
longer than the normal mode. In the PSG model the drifting periodicity is 
inversely proportional to the screening factor in the IAR \citep{MBM20,BMM20b}.
Hence, an increase in the drifting periodicity requires changes in the 
screening condition, which are governed by several factors including the 
surface magnetic field strength. On the other hand the phase behaviour of the 
slow-drift mode closely follow the normal mode, suggesting that the relative 
orientation of the surface magnetic fields with respect to the LOS remains 
unchanged in the two emission modes. Thus the surface magnetic field strength 
changes without affecting its relative orientation, and needs to be accounted 
for while modelling the drifting behaviour.

\section*{Acknowledgements}
This work was partially supported by the grant 2020/37/B/ST9/02215 of the 
National Science Centre, Poland. We would like to thank the Ministry of 
Education and Science of Poland for granting funds for the Polish contribution 
to the International LOFAR Telescope, LOFAR2.0 upgrade (decision number: 
2021/WK/2) and for maintenance of the LOFAR PL-610 Bor\'owiec, LOFAR PL-611 
\L{}azy, LOFAR PL-612 Ba\l{}dy, stations (decision numbers: 
30/530252/SPUB/SP/2022, 29/530358/SPUB/SP/2022, 28/530020/SPUB/SP/2022, 
respectively).

\section*{Data Availability}
The data underlying this article will be shared on reasonable request to the 
corresponding author.

\bibliographystyle{mnras}
\bibliography{reflist} 

\appendix
\section{Emission Mode sequence} \label{sec:app}

A detailed list of pulse sequence in each emission mode is reported in 
Table.\ref{tab:modelist}. We also show the longer null intervals within the 
slow-drift mode.

\begin{table*}
\caption{Emission Mode sequence}
\label{tab:modelist}
\begin{tabular}{lccccc@{\hspace{7\tabcolsep}}lcccc}
\hline
  \multicolumn{5}{c}{25 June, 2019} & & \multicolumn{5}{c}{7 October, 2022} \\
\hline
 Mode & Pulse Range & Length & Null & Length &  & Mode & Pulse Range & Length & Null & Length \\
   & ($P$) & ($P$) & ($P$) & ($P$) &  &  & ($P$) & ($P$) & ($P$) & ($P$) \\
\hline
   &  &  &  &  &  &  &  &  &  &  \\
 Normal & 1 - 679 & 679 &  &  &  & Normal & 1 - 1573 & 1573 &  &  \\
   &  &  &  &  &  &  &  &  &  &  \\
 Slow-drift & 680 - 711 & 32 & 684 - 691 & 8 &  & Slow-drift & 1574 - 1606 & 33 & 1583 - 1592 & 10 \\
   &  &  &  &  &  &  &  &  &  &  \\
 Normal & 712 - 1137 & 426 &  &  &  & Normal & 1607 - 2302 & 696 &  &  \\
   &  &  &  &  &  &  &  &  &  &  \\
 Slow-drift & 1138 - 1165 & 28 & 1142 - 1146 & 5 &  & Slow-drift & 2303 - 2349 & 47 & 2316 - 2324 & 9 \\
   &  &  &  &  &  &  &  &  &  &  \\
 Normal & 1166 - 1769 & 604 &  &  &  & Normal & 2350 - 3906 & 1557 &  &  \\
   &  &  &  &  &  &  &  &  &  &  \\
 Slow-drift & 1770 - 1806 & 37 & 1778 - 1783 & 6 &  & Slow-drift & 3907 - 3946 & 40 & 3914 - 3923 & 10 \\
   &  &  &  &  &  &  &  &  &  &  \\
 Normal & 1807 - 1941 & 135 &  &  &  & Normal & 3947 - 7258 & 3312 &  &  \\
   &  &  &  &  &  &  &  &  &  &  \\
 Slow-drift & 1942 - 1974 & 33 & 1946 - 1949 & 4 &  & Slow-drift & 7259 - 7293 & 35 & 7266 - 7271 & 6 \\
   &  &  &  &  &  &  &  &  &  &  \\
 Normal & 1975 - 3691 & 1717 &  &  &  & Normal & 7294 - 7957 & 664 &  &  \\
   &  &  &  &  &  &  &  &  &  &  \\
 Slow-drift & 3692 - 3719 & 28 & 3704 - 3707 & 4 &  & Slow-drift & 7958 - 8005 & 48 & 7963 - 7967 & 5 \\
   &  &  &  &  &  &  &  &  &  &  \\
 Normal & 3720 - 4345 & 626 &  &  &  & Normal & 8006 - 9089 & 1084 &  &  \\
   &  &  &  &  &  &  &  &  &  &  \\
 Slow-drift & 4346 - 4386 & 41 & 4352 - 4355 & 4 &  & Slow-drift & 9090 - 9122 & 33 & 9097 - 9103 & 7 \\
   &  &  &  &  &  &  &  &  &  &  \\
 Normal & 4387 - 4472 & 86 &  &  &  & Normal & 9123 - 9246 & 124 &  &  \\
   &  &  &  &  &  &  &  &  &  &  \\
 Slow-drift & 4473 - 4506 & 34 & 4479 - 4484 & 6 &  & Slow-drift & 9247 - 9279 & 33 & 9252 - 9256 & 5 \\
   &  &  &  &  &  &  &  &  &  &  \\
 Normal & 4507 - 4678 & 172 &  &  &  & Normal & 9280 - 9356 & 78 &  &  \\
   &  &  &  &  &  &  &  &  &  &  \\
 Slow-drift & 4679 - 4713 & 35 & 4686 - 4689 & 4 &  & Slow-drift & 9357 - 9378 & 22 & 9360 - 9364 & 5 \\
   &  &  &  &  &  &  &  &  &  &  \\
 Normal & 4714 - 5092 & 379 &  &  &  & Normal & 9379 - 12252 & 2874 &  &  \\
   &  &  &  &  &  &  &  &  &  &  \\
 Slow-drift & 5093 - 5116 & 24 & 5099 - 5101 & 3 &  & Slow-drift & 12253 - 12324 & 72 & 12263 - 12273 & 11 \\
   &  &  &  &  &  &  &  &  &  &  \\
 Normal & 5117 - 5659 & 543 &  &  &  & Normal & 12325 - 12355 & 31 &  &  \\
   &  &  &  &  &  &  &  &  &  &  \\
 Slow-drift & 5660 - 5686 & 27 & 5665 - 5670 & 6 &  & Slow-drift & 12356 - 12441 & 86 & --- & 0 \\
   &  &  &  &  &  &  &  &  &  &  \\
 Normal & 5687 - 6603 & 917 &  &  &  & Normal & 12442 - 12992 & 551 &  &  \\
   &  &  &  &  &  &  &  &  &  &  \\
 Slow-drift & 6604 - 6640 & 37 & 6609 - 6614 & 6 &  & Slow-drift & 12993 - 13025 & 33 & 13001 - 13006 & 6 \\
   &  &  &  &  &  &  &  &  &  &  \\
 Normal & 6641 - 6857 & 217 &  &  &  & Normal & 13026 - 14448 & 1423 &  &  \\
   &  &  &  &  &  &  &  &  &  &  \\
 Slow-drift & 6858 - 6901 & 44 & 6863 - 6871 & 9 &  & Slow-drift & 14449 - 14492 & 44 & 14458 - 14465 & 8 \\
   &  &  &  &  &  &  &  &  &  &  \\
 Normal & 6902 - 8116 & 1215 &  &  &  & Normal & 14493 - 15080 & 588 &  &  \\
   &  &  &  &  &  &  &  &  &  &  \\
\hline
\end{tabular}
\end{table*}

\bsp	
\label{lastpage}
\end{document}